\documentclass[conference]{IEEEtran}
\usepackage[hidelinks]{hyperref}
\ifCLASSINFOpdf
\else
\fi
\usepackage{url}

\begin{document}
%
\title{The BioDynaMo Project: Creating a Platform for Large-Scale Reproducible Biological Simulations}


\author{
\IEEEauthorblockN{
Lukas Breitwieser\IEEEauthorrefmark{1}, 
Roman Bauer\IEEEauthorrefmark{2},
Alberto Di Meglio\IEEEauthorrefmark{1},
Leonard Johard\IEEEauthorrefmark{3},\\
Marcus Kaiser\IEEEauthorrefmark{2},
Marco Manca\IEEEauthorrefmark{1},
Manuel Mazzara\IEEEauthorrefmark{3},
Fons Rademakers\IEEEauthorrefmark{1},
Max Talanov\IEEEauthorrefmark{4},
}
\IEEEauthorblockA{
Affiliation: 
\IEEEauthorrefmark{1}CERN (Switzerland),
\IEEEauthorrefmark{2}Newcastle University (United Kingdom),\\
\IEEEauthorrefmark{3}Innopolis University (Russian Federation), \IEEEauthorrefmark{4}Kazan Federal University  (Russian Federation)\\
Email: biodynamo-talk@cern.ch
}
}


%


\IEEEoverridecommandlockouts
\IEEEpubid{\makebox[\columnwidth]{This work is licensed under a \href{https://creativecommons.org/licenses/by/4.0/}{CC-BY-4.0 license}.\hfill} \hspace{\columnsep}\makebox[\columnwidth]{ }}

\maketitle

\begin{abstract}

Computer simulations have become a very powerful tool for scientific research. In order to facilitate research in computational biology, the BioDynaMo project aims at a general platform for biological computer simulations, which should be executable on hybrid cloud computing systems. This paper describes challenges and lessons learnt during the early stages of the software development process, in the context of implementation issues and the international nature of the collaboration.

\end{abstract}


%
\IEEEpeerreviewmaketitle

\section{Introduction}

The BioDynaMo project is a long term effort in the field of biological simulation to build a scalable and flexible platform. The purpose is to give life scientists access to increasing amounts of computational resources and provide a framework that hides the computational complexity, allows them to focus on their research, and promotes reusability and reproducibility of the results from shared open access data. 
In order to have an impact on the community, the system must be flexible enough to execute simulations from different specialities with possibly quite distinct requirements. 
  



The project started as a code modernization initiative, inspired by the scientific principles underlying the simulation software Cx3D \cite{zubler}.
Cx3D is a software framework that is able to simulate the development of neural tissue, based on physical mechanisms and neural growth \cite{zubler}. However, Cx3D can not leverage cloud computing systems or coprocessors, and so is limited in terms of the simulation size and complexity. Moreover, Cx3D is limited in terms of extendability and modifiability for other purposes in computational biology.

In general, software modernization is a collective term that subsumes a variety of activities. In our case it means transforming the application from Java to C++ and changing the architecture in a way to utilize multiple levels of parallelism offered by today's hardware and modern distributed computing models.

\section{Software Development Practices}

Although Cx3D has a very compact code base (15 kLOC), it is able to perform complex simulations like ``cortical lamination''. However, the absence of modern software development practices such as automated tests, continuous integration, coding standards compliance, and code reviews hinders a sustainable development process. 

Our first step was to introduce development techniques and infrastructure aimed at improving code quality and maintainability, which are essential for our long time effort. Based on available effort, we opted for testing the whole application rather than writing unit tests for the entire codebase. Existing demo simulations were taken and transformed into test cases. 
The resulting simulation state is then transformed into JSON format and compared to a ground truth obtained from Cx3D 0.03. A public code repository was created on Github \cite{bdm_github}
and connected to the continuous integration service Travis-CI \cite{Travi75:online}
that automatically checks every code change if it generates the correct results. This procedure proved to be a good choice given the goal of improving application performance without changing the final output. 
Furthermore, a coding styleguide was selected to ensure that code is readable, maintainable and follows best practices. A coding standard is only helpful if it is followed by the developers. Thus, tools are needed that help to conform to these rules. We chose the Google C++ styleguide \cite{google_style_guide}
which comes with an Eclipse code formatting definition and cpplint, a tool that checks code for violations.
A dedicated ``BioDynaMo Developers Guide'' \cite{bdm_dev_guide}
introduces new developers to the project, describes conventions beyond the coding style, e.g. usage of the revision system git, and stresses the importance of testing and documentation. External contributions are introduced through Github's pull request system and are reviewed before they are merged into our repository. Github also offers an issue tracking system that is helpful to report and document software errors and to plan future work packages. Moreover, communication is an important aspect especially with project partners based in different countries. Our team uses the message system Slack \cite{slack} for real time low bandwith communication which integrates well with Github and Travis-CI. Alternatively, we have set-up two mailing lists for asynchronous communication. Additionally, conference calls using Skype and periodic plenary meetings complement our communication toolbox and help us to coordinate this project.




\section{Modernizing Legacy Code: Examples of the Methodologies Applied}
High performance and high scalability are the prerequisites to address ambitious research questions like modeling epilepsy. Our efforts in code modernization were driven by the goal to remove unnecessary overhead and update the software design to tap the unused potential enabled by the paradigm shift to multi and many-core distributed systems. 

The Intel Modern Code Development Challenge organized in 2015 with CERN and Newcastle focused on optimizing sequential C++ brain simulation code provided by the Newcastle University in the UK. 
The contest followed a gamification approach where participating students competed against each other to win an internship at CERN. The ranking was based on the runtime of the provided simulation.
Using data layout transformations (array of structures to structure of arrays), parallelization with OpenMP, a custom memory allocator and Intel Cilk Plus array notation, the winner was able to decrease the runtime by a factor of 320. 
This clearly shows the economic potential of code modernization efforts coupled with gamification and encourages to repeat the challenge.

Furthermore, we ported the Java code base to C++. This language is better suited for high performance computing as it is compiled to native machine code removing the overhead of running in a virtual machine and provides the right ecosystem for parallelization and optimization. The following iterative porting approach has been chosen. First, a Java class is selected and replaced by its C++ translation. In the second step, this C++ class is connected to the remaining Java application. Finally, the Java/C++ hybrid is compiled and used to execute a number of tests. If all tests pass, the developer can proceed with the next iteration by selecting another Java class. 
On the other hand this means that errors, indicated by test failures, must have been introduced by code changes since the last iteration. Therefore, this procedure significantly simplifies debugging.
Although this approach is associated with additional development overhead in connecting classes in C++ to Java, it gives the benefit of obtaining a runnable system after each iteration. Without that additional effort, the first time the C++ version would be able to execute tests, would be at the very end, after all classes have been ported. 
Porting would have been a lot easier if every class / function had sufficient unit tests. In this scenario connecting both languages would no longer be required since tests could be executed for each function independently. Testing the whole simulation software had another drawback: floating point differences on diverse systems amplified over many iterations and were responsible for test failures although the code was correct. We fixed that issue using the math-library \texttt{crlibm}
to obtain reproducible results across different environments as suggested in \cite{mcintosh2006massive}. Setting up the whole development environment and porting the application took six months.
A preliminary performance benchmark of the single threaded, non vectorized C++ version showed promising performance improvements of up to 4.8x with a median of 1.7x.

\section{Conclusion}
The field of computational biology covers a wide range of scientific topics, each producing many different scientific models, such as for instance described by \cite{bauer2014developmental}, \cite{freund2014numerical} and \cite{izhikevichEdelman2008large}. Hence, a general platform for biological research should be able to meet a number of different requirements. It is crucial that this diversity of the prospective users is already taken into account during the software development process. Incorporating such diversity means that the multidisciplinary project team of BioDynaMo must be able to efficiently interact, and make decisions based on the expertise of each team member.

In addition to these more scientifically-centered aspects, also considerable challenges arise from a computational/technological point of view. First steps towards such efficient software implementation have been made in the context of the ``Intel Modern Code Developer Challenge'' competition. Overall, we believe we have created a collaborative foundation for the efficient continuation of the very ambitious software development project of BioDynaMo.

However, considerable challenges remain in the current software development process. The verification and validation of the software is paramount. The recent study of \cite{Eklundetal2016cluster} demonstrates the extraordinary risks that arise when the correctness and validity of software tools for scientific research are not properly assessed. Moreover, the efficient communication and orchestration among the members are crucial components of this international project. We have identified these key aspects to require further efforts in parallel to the overall development process.

\section*{Acknowledgment}

This work was possible thanks to the support by CERN and CERN openlab “Code Modernization” program in cooperation with Intel; by the Human Green Brain Project (\url{www.greenbrainproject.org}) through the Engineering and Physical Sciences Research Council (EP/K026992/1); by Innopolis University, and its Service Science and Engineering lab (SSE); and by SCImPULSE Foundation. The funding institutions had no role in the design of the project, decision to publish, or preparation of the manuscript.



%

\bibliographystyle{IEEEtran}
\bibliography{wssspe4}

\end{document}